**Magnetic flux ropes within reconnection exhausts close to the centers of heliospheric current sheets near the Sun**

Dae-Young Lee[1], Dooyoung Choi[1], Kyung-Eun Choi[2], Sung Jun Noh[3]

[1]Department of Astronomy and Space Science,
Chungbuk National University, Chungbuk 28644, Korea;
[2]Space Sciences Laboratory, University of California, Berkeley, CA, USA;
[3]ISR-1: Space Science and Applications, Los Alamos National Laboratory, NM, USA

**Abstract**

Understanding the relationship between magnetic flux ropes and magnetic reconnection is fundamental to both space and astrophysical plasma studies. In this study, we report on two consecutive heliospheric current sheet (HCS) crossings by Parker Solar Probe (PSP), separated by ~10.5 hours, at a heliocentric distance of ~12 solar radii. For each crossing, we identified a series of flux ropes embedded within reconnection exhausts on the sunward side of X-line. Their passage durations are <20sec, corresponding to spatial scales of a few thousands kilometers, still larger by three orders of magnitude than ion inertial length. This identification was possible particularly during intervals when PSP was closest to the HCS center. These flux ropes are distinguishable from the background exhausts by enhancements in magnetic field strength, significantly in the guide field component, travel speed slightly faster (typically by <10km/s) than surrounding outflows, and often accompanied by, though not always, increased density and reduced temperature. We attribute their origin to secondary reconnection within the exhausts and subsequent merging of smaller flux ropes into larger structures, consistent with predictions by various simulations. We suggest that such flux ropes are most readily identifiable at the HCS center where the background magnetic field is weakest so that the relative enhancement in flux rope field becomes most prominent. This observational advantage is particularly notable closer to the Sun where the high ambient magnetic field strength can otherwise obscure such structures unless the spacecraft trajectory remains within the HCS central region for a sufficient duration.

# 1. Introduction

Magnetic flux ropes or flux tubes (terms used interchangeably throughout this paper) of various sizes are widely recognized as important structures in space and astrophysical plasmas, with significant implications for energy transport, particle acceleration, and magnetic topology. Their origin and evolution are of particular interest across diverse environments, including planetary magnetospheres, the solar corona, and the heliosphere. The launch of the Parker Solar Probe (PSP) has enabled unprecedented access to the near-Sun solar wind, allowing the identification of flux ropes closer to the Sun than ever before. Recent studies using PSP data have reported flux ropes with durations ranging from several minutes to multiple days at heliocentric distances of ~0.1–0.3 AU (e.g., J. F. Drake et al. 2021; L. L. Zhao et al. 2020, 2021; Y. Chen & Q. Hu 2022; Y. Chen et al. 2020, 2021, 2023).

The relationship between flux ropes and magnetic reconnection, which is a fundamental plasma process with broad relevance in both space and astrophysical contexts, is an active research topic. Reconnection is widely regarded as a viable mechanism for the generation of small-scale magnetic flux ropes both in the solar corona (J. F. Drake et al. 2021; B. Lavraud et al. 2020; Réville et al. 2020, 2022) and in the solar wind, particularly in the vicinity of the heliospheric current sheet (HCS) (M. B. Moldwin et al. 2000; M. L. Cartwright & M. B. Moldwin 2008; A. K. Higginson & B. J. Lynch 2018; E. Sanchez-Diaz et al. 2019; B. Lavraud et al. 2020).

For instance, simulations have demonstrated the formation and evolution of flux ropes produced by reconnection at the top of the helmet streamer belt within 30 $R_\odot$ (e.g., A. K. Higginson & B. J. Lynch 2018). Several PSP observations have shown density blobs and flux ropes released from the tips of helmet streamers close to the Sun (B. Lavraud et al. 2020; K.-E. Choi et al. 2024; P. Liewer et al. 2024), all attributed to reconnections.

Indeed, reconnection exhausts during a crossing of HCSs have long been reported at 1 AU (J. T. Gosling et al. 2005; B. Lavraud et al. 2009), and more recent studies have revealed that such signatures are far more frequent closer to the Sun (T. Phan et al. 2020, 2021, 2024). These reconnection exhausts are often identified by characteristics such as plasma flow jets in association with electron strahl, and bifurcated magnetic field structure.

Notably, even subscale bifurcated current sheets – ranging in thickness of ~20-2000 ion inertial length - with flow jets have been identified within the HCS exhaust close to the Sun, implying secondary reconnections within the exhaust of the primary reconnection (T. Phan et al. 2024). Secondary reconnection has been widely considered a major mechanism for generating flux ropes of a small scale, which subsequently undergo merging with one another (W. H. Mattaeus & S. L. Lamkin 1986; J. F. Drake et al. 2006; S. Servidio et al. 2009; G. Lapenta et al. 2015; L. Comisso et al. 2016; C. Dong et al. 2018; H. Arro et al. 2020; H. Arnold et al. 2021; S. Eriksson et al. 2022; M. I. Desai et al. 2025).

In the present study, we report on two consecutive HCS crossings, each revealing a series of small-scale (still much larger than ion inertial length) magnetic flux ropes identified within reconnection exhausts, particularly during intervals when PSP was closest to the center of the HCS. These observations were made over a limited time window when PSP was retrograding in the Sun's rotating frame near the Sun at a heliocentric distance of ~12 $R_\odot$, near perihelion of Encounter #17. This period coincided with solar maximum conditions, during which the HCS was highly warped, observed as two closely spaced HCS crossings by PSP.

The paper is organized as follows. Section 2 provides an overview of the two consecutive HCS crossing events. Section 3 presents detailed observations of a series of flux ropes embedded within the reconnection exhausts near the HCS center. Discussion and conclusions are provided in Section 4.

## 2. Overview of two consecutive HCS crossings

PSP encountered two HCSs on 2023 September 27 and 28, at heliocentric distances of ~11.6 $R_\odot$ and ~12.1 $R_\odot$, respectively. These crossings were separated by ~10.5 hours, one occurring several hours before perihelion and the other several hours after.

Figure 1(a) schematically illustrates the geometry of these crossings through a highly warped current sheet, based on the corresponding Potential Field Source Surface (PFSS) model map obtained from the Wilcox Solar Observatory (J. T. Hoeksema et al. 1983) shown in Figure 1(b). The PFSS solution indicates that the two HCS structures encountered by PSP were separated by ~40° in Carrington longitude. During the first crossing event on September 27, PSP's motion was primarily in the tangential (T) direction with a speed of $V_T$=173km/s, and slightly sunward at $V_R$=-24 km/s in the Radial-Tangential-Normal (RTN) coordinate system. At the second crossing event on September 28, PSP's motion was similar, with $V_T$=164km/s and slightly anti-sunward component of $V_R$=43 km/s. During both crossings, PSP at these speeds was retrograding in the Carrington rotation frame.

The PFSS map further shows that the two HCS planes encountered by PSP were highly inclined relative to the heliographic equator, consistent with the complex magnetic topology expected during solar maximum. It suggests that the HCS planes were tilted somewhat relative to the R-N plane toward the T-direction.

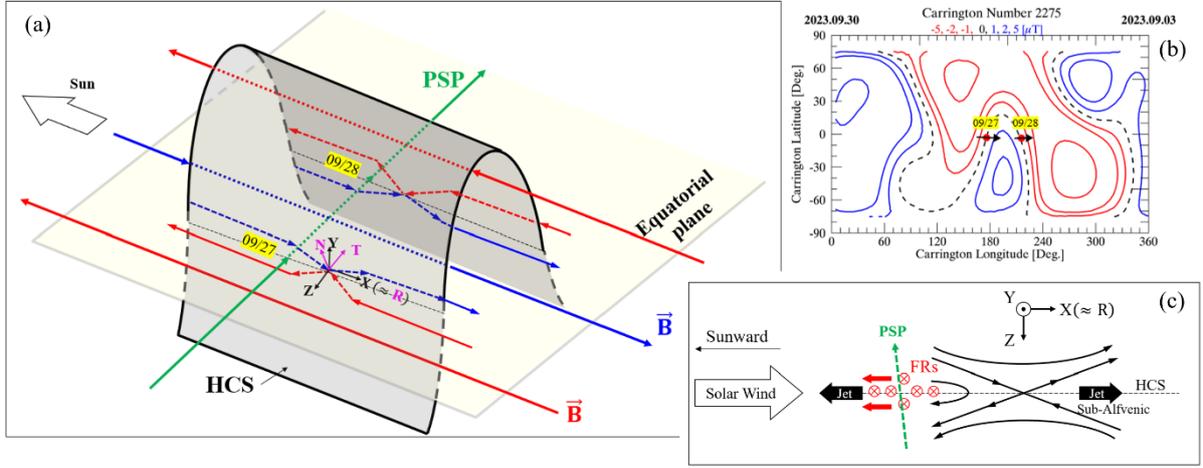

Figure 1. Highly warped heliospheric current sheet (HCS) and associated magnetic reconnection. (a) Schematic illustration of two consecutive crossings of the warped HCS by Parker Solar Probe (PSP) at r ≈ 12$R_\odot$ on September 27 and 28, 2023, along with associated magnetic reconnection sites. (b) Potential Field Source Surface (PFSS) model map of the coronal magnetic field at r=2.5$R_\odot$, showing the projected PSP locations at the times of the two HCS crossings. (c) Schematic diagram of magnetic reconnection at the HCS, showing magnetic flux ropes (FRs) embedded within the reconnection exhaust jet. The local current sheet coordinate system (X, Y, Z) is indicated, in contrast to the conventional radial–tangential–normal (R, T, N) system.

In the following section, we describe in detail the occurrence of magnetic reconnection at both HCS crossings, with particular emphasis on the series of magnetic flux ropes embedded within the reconnection exhausts (as schematically depicted in Figure 1(c)). For the analysis, we use in situ observations from PSP. The PSP FIELDS instrument suite (S. D. Bale et al. 2016) provides DC magnetic field measurements from fluxgate magnetometers (MAG). The PSP SWEAP instrument suite (J. C. Kasper et al. 2016) provides the proton density, velocity, and temperature from the SPAN-ion instrument (R. Livi et al. 2020) and the pitch angle distribution of suprathermal electrons from the Solar Probe Analyzer Electron (SPAN-E) instrument (P. L. Whittlesey et al. 2020).

## 3. Series of flux ropes embedded within reconnection exhausts
### 3.1 Case on 2023 September 27

The data relevant to the HCS crossing on this date are presented in Figure 2 where we have identified the full HCS crossing interval over ~8 mins (two dashed vertical lines). This interval is characterized by well-known features including (i) the reversal of electron strahl direction from 180° before this interval to 0° after it (Figure 2(a)), (ii) the overall notable decrease of the magnetic field magnitude despite intermittent recoveries to near pre-crossing levels (Figure 2(b)), and (iii) the complete polarity reversal in the asymptotic values of the radial component of the magnetic field, $B_R$, shifting from ~-650nT to ~+650nT, before and after this interval, respectively (Figure 2(c)) consistent with the strahl direction change.

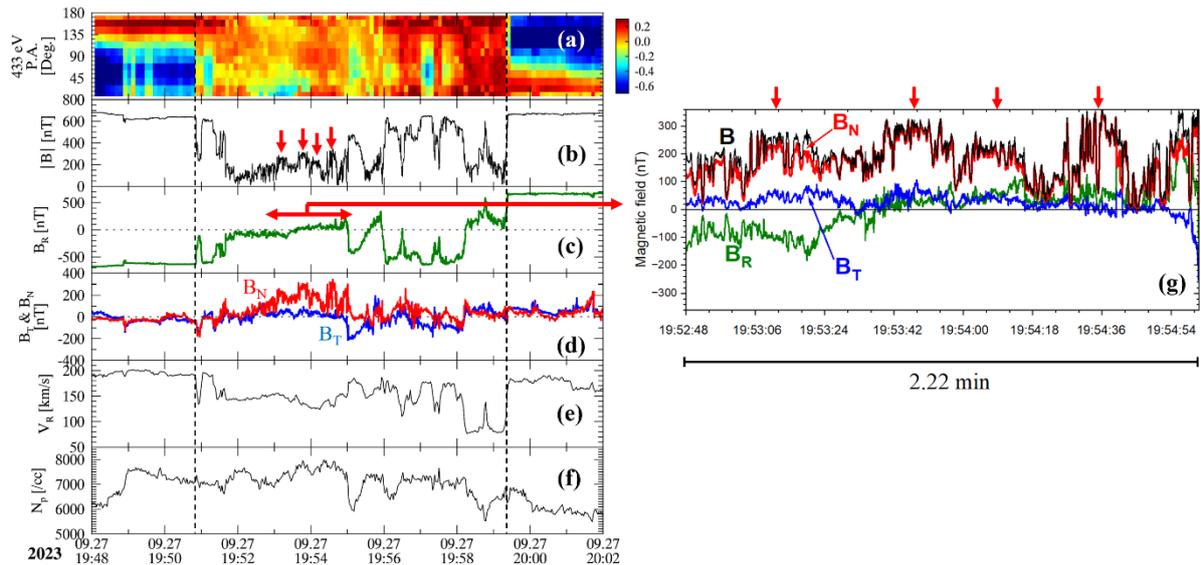

Figure 2. Observations of the heliospheric current sheet (HCS) crossing on September 27, 2023. (a) Pitch-angle distribution of 433 eV suprathermal electrons. (b) Magnetic field magnitude |B|, with red arrows marking four enhancement events. (c) Radial magnetic field component $B_R$, with the

horizontal red arrow indicating the interval of main interest. (d) Tangential ($B_T$, blue) and normal ($B_N$, red) magnetic field components. (e) Radial component of the solar wind velocity $V_R$. (f) Proton number density $N_p$. (g) Expanded view of the magnetic field data during the ~2.2-minute interval marked in panel (c), showing the four enhancement events in detail.

Evidence for magnetic reconnection at this HCS can be drawn as follows. First, during the identified HCS crossing interval, the radial flow $V_R$ in Figure 2(e) exhibits an overall reduction, occurring intermittently and most prominently when PSP was passing close to the HCS center. In addition, Figure 2(a) indicates the existence of overall counter-streaming electron strahl throughout the full crossing time (despite the simultaneous existence of increased electron fluxes at all pitch angles). The drop in $V_R$ in the Sun's frame in combination with the bi-directional strahl is a well-established signature of a reconnection outflow jet when PSP traverses the sunward side of a reconnection X-line (J. T. Gosling et al. 2006; T. Phan et al. 2020, 2021).

We stress that the magnetic field intensity (Figure 2(b)) is weakest for a several-min interval starting from ~19:52 UT, during which the magnitude of $B_R$ is also smallest (Figure 2(c)), implying the closest approach of PSP to the HCS center. At other times, PSP's position varied relative to the HCS center, and it was often located much away from the HCS center.

We draw particular attention to four instances of enhanced magnetic field strength (marked by red vertical arrows in Figure 2(b)), occurring within a ~2.2-minute interval (marked by the horizontal red arrowed line in Figure 2(c)). These events are notably characterized by distinct increases in the normal component of the magnetic field, $B_N$ (the red line in Figure 2(d)). This feature is more clearly illustrated in Figure 2(g), which zooms in on the 2.2 min interval and highlights the four B enhancement events (red vertical arrows). The B enhancements range from ~40% to ~50% relative to the neighboring field magnitude.

Now we focus on and examine the detailed features during the ~2.2 min interval within which the four instances of enhanced B are identified. The relevant data are shown in Figure 3 where the four events are highlighted based on their B enhancements. Each event lasts ~12–18 seconds corresponding to spatial scales of several thousand kilometers (~$10^{-5}$AU ≈ ~$10^{-3} R_\odot$), which are significantly larger (by three orders of magnitude) than the local proton inertial length of ~2.6 km. The four events are separated by a few to <20 seconds from one another.

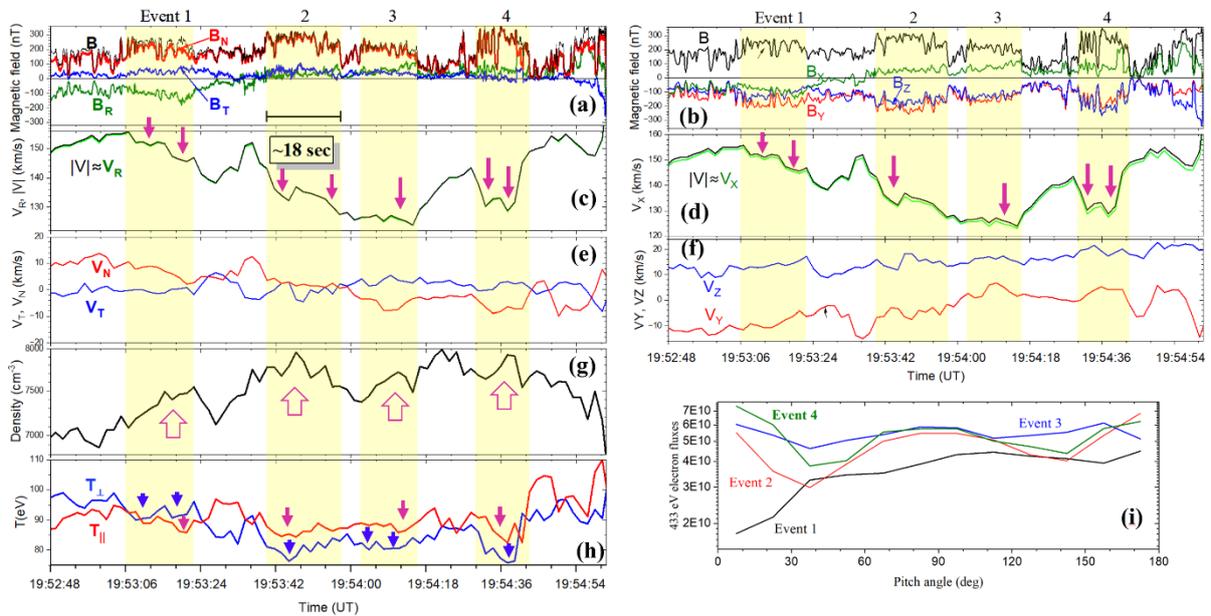

Figure 3. Detailed observations for the ~2.2-minute interval highlighted in Figure 2. (a) Magnetic field components in RTN coordinates. (b) Magnetic field components in the local current sheet coordinate system (X, Y, Z). (c) The magnitude of the solar wind bulk velocity |V| and its radial component $V_R$ in RTN coordinates. (d) $V_X$ in the current sheet coordinates. (e) Tangential and normal velocity components $V_T$ and $V_N$ in RTN. (f) $V_Y$ and $V_Z$ in the current sheet coordinates. (g) Proton number density. (h) Parallel and perpendicular components of the proton temperature ($T_\parallel$ and $T_\perp$). (i) Average suprathermal electron flux (433 eV) vs. pitch angle during each of the four events (highlighted yellow).

Several important features are evident. First, the B enhancements are dominated by increases in $B_N$ (red in Figure 3(a)). Following the method of S. Eriksson et al. (2022), we transformed the plasma flow and magnetic field observations from the RTN coordinate system into a local current sheet coordinate system (X, Y, Z), as depicted in Figure 1(c). The normal vector of the current sheet is given by [-0.02, 0.65, -0.76] in the RTN coordinates, representing the orientation and tilt of the current sheet relative to the RTN axes. In this current sheet coordinate system, $B_X$ corresponds to the component of the magnetic field that reconnects across the HCS (aligned with the reconnection exhaust direction), $B_Y$ corresponds to the guide field component, and $B_Z$ is the component normal to the current sheet. Figure 3(b) displays the magnetic field in this transformed coordinate system, indicating that the B enhancements are caused significantly by the magnitude increases of $B_Y$ (red), rising from the surrounding values of ~60-100nT to peak values of ~160-240nT, and by similar increases of $B_Z$ magnitude (blue) together. In contrast, $|B_X|$ (olive) remains comparatively small throughout most of the interval. These observations suggest the enhanced magnetic field vectors are inclined by ~45° primarily relative to the current sheet plane (X-Y plane). Second, each of the four events is associated with a further decrease in $|V|$ (vertical magenta arrows in Figure 3(c) and 3(d)), which is largely governed by $V_R$ (or equivalently $V_X$ in the current sheet coordinate system). The magnitude of this subscale velocity drop is slightly over 10km/s for Event 4 and several km/s for the other three events. Lastly, each magnetic enhancement is accompanied by an increase in plasma density (Figure 3(g)) and a decrease in temperature (Figure 3(h)), as indicated by vertical arrows in the corresponding panels. The temperature decrease manifests mostly in both components of the temperature tensor, although it is overall less pronounced than the associated density increases.

These features lead us to interpret the four events as distinct "flux tubes" embedded within the reconnection exhaust, moving sunward (in the ambient solar wind frame) at speeds slightly greater than those of the surrounding sunward outflows of the exhaust. Their differing propagation speeds may cause some velocity shear between the flux tubes and the surrounding outflows. The solar connectivity of the flux tubes can be inferred from the suprathermal electron data (e.g., K.-E. Choi et a. 2019). Figure 3(i) indicates the existence of counter-streaming strahl for three events except for Event 1, implying the possibility of closed field line topology for the three events. The orientation of these magnetic flux tubes is not aligned with the primary background magnetic field direction, which is predominantly radial. Instead, the flux tubes exhibit highly inclined orientations relative to the radial direction, a feature that has been noted in previous studies (e.g., A.K. Higginson & B.J. Lynch 2018; K.-E. Choi et al. 2022).

At present, it is not entirely clear whether these structures are typical "flux ropes" which are usually characterized by helically wound magnetic field lines. A defining feature of in situ flux rope observations is a smooth rotation in one or more magnetic field components; however, such rotational signatures are absent in the first three events and are only weakly suggested by a sign change in $B_T$ during Event 4. Nevertheless, we do not rule out the possibility that PSP traversed only a limited portion of each flux rope cross-section, potentially missing the full rotation or sign change in the magnetic field, while each flux ropes is largely dominated by the axial field. In this paper we do not attempt to distinguish flux rope and flux tube rigorously from a physical standpoint.

Another interesting feature is the presence of large-amplitude compressive oscillations observed in all three magnetic field components at a low frequency of ~0.2 Hz, much lower than the local proton gyrofrequency for B = 250 nT (~3.8 Hz). Notably, these waves persist throughout the entire interval shown in Figure 3, including during each of the flux tube events.

### 3.2 Case on 2023 September 28

The relevant data for the September 28 crossing are shown in Figure 4 in the same format as Figure 2. Based on the electron strahl observations, we identify the full crossing interval of this HCS (indicated by vertical dashed lines) as the period during which the strahl direction reverses from 0° before this interval to 180° after it (Figure 4(a)) while the magnetic field magnitude becomes weak (Figure 4(b)). Additionally, the sharp changes in $B_R$ (Figure 4(c)) near the two edges of this interval and the plateau in between imply that the current sheet was bifurcated, a feature commonly observed in reconnecting current sheets (e.g., T. Phan et al. 2020).

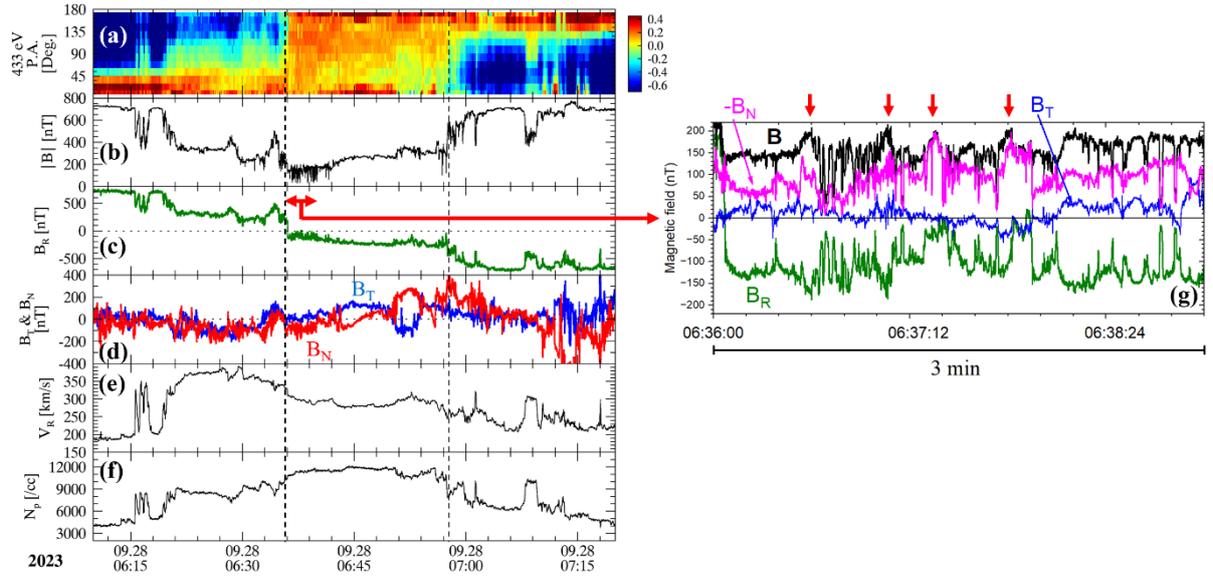

Figure 4. Observations of the heliospheric current sheet (HCS) crossing on September 28, 2023, in the same format as Figure 2. Note that panel (g) shows an expanded view of the magnetic field data over a 3-minute interval, highlighting four B enhancement events and the significance of the normal component. In panel (g), $-B_N$ (magenta) is shown for easier comparison with the total B (black).

Similar to the September 27 event, evidence for magnetic reconnection at this HCS is provided by the reduction in the radial component of the solar wind speed, $V_R$ (Figure 4(e)). The drop in $V_R$, along with the presence of overall counter-streaming strahl suggests that PSP crossed the sunward side of a reconnection X-line, similar to the event observed on 2023 September 27.

We focus in particular on a few minute interval (horizontal arrowed line in Figure 4(c)), during which the magnetic field magnitude is at its lowest (Figure 4(b)), that of $B_R$ is also very small (Figure 4(c)), and magnetic field fluctuations are most prominent. Figure 4(g) is a zoomed-in view of the magnetic field data for this short interval. Despite the large fluctuations, we identify four intervals of enhanced B (red vertical arrows), similar to the case observed during the September 27 event in the previous section.

Figure 5 shows the relevant data for this short interval in the same format as Figure 3. These flux tubes are characterized by enhanced B by ~40% relative to the neighboring field magnitude. Each event lasts for ~4 to 12 seconds, corresponding to spatial scales of a few thousands kilometers (far larger than the ion inertial length of ~2.1km). The separation between adjacent events ranges from ~6 to 24 seconds.

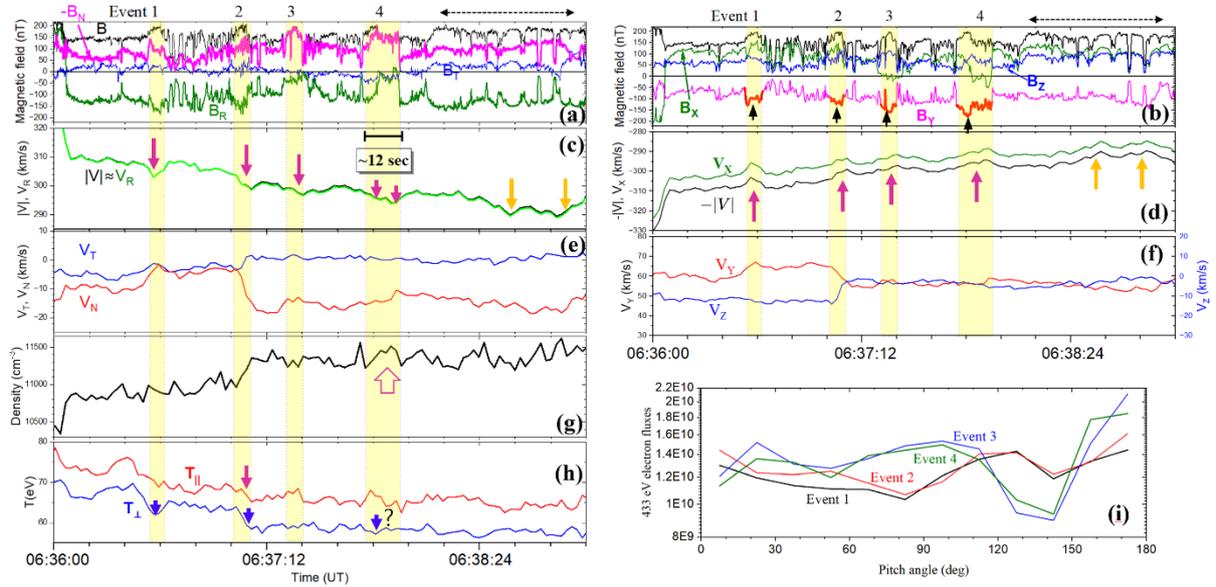

Figure 5. Detailed observations for the 3-minute interval highlighted in Figure 4 in the same format as Figure 3. Note that $-B_N$ (magenta) is shown in panel (a) and $-|V|$ (black) in panel (d) for easier comparison with other parameters. In panel (b), the enhanced portions in $B_Y$ magnitude are highlighted by thick red segments and black arrows.

Key features associated with these flux tubes are similar to those found in the September 27 case, although some are less obvious in this crossing. In particular, two features most clearly distinguish the flux ropes from the surroundings. First, the B enhancements are caused by increases in $B_N$ dominantly over the other two components for Events 3 and 4, and to a lesser extent but still noticeably for Events 1 and 2 (magenta for $-B_N$ in Figure 5(a)). In the current sheet coordinate system (determined by a normal vector of [-0.04, 0.81, -0.59] in RTN coordinates), this trend is similarly reflected in enhancements of $|B_Y|$, with the enhanced portions in $|B_Y|$ highlighted by thick red segments and black arrows in Figure 5(b). The average increase in $|B_Y|$ during each event, relative to the surrounding average $|B_Y|$ (calculated over a 24-sec window around each event) ranges from ~22.6% (Event 2) to ~53.5% (Event 1). Second, each of the four events are associated with a further decrease in $|V|$ (vertical red arrows in Figure 5(c)) primarily driven by changes in $V_R$. This is equivalently, and more clearly, reflected as the increase in $V_X$ in the local current sheet coordinate system (vertical red arrows in Figure 5(d)). The magnitude of these velocity changes is less than 10km/s.

On the other hand, the association with density variations is not consistent across all events (Figure 5(g)): a density enhancement may be observed for Event 4, while clear variations are not apparent for the other three flux tubes. Figure 5(h) indicates that a decrease in temperature in at least one component of the temperature tensor may be identifiable for Events 1 and 2, but not for Events 3 and 4. Additionally, these flux tubes are mostly associated with counter-streaming strahls as shown in Figure 5(i). Lastly, these flux rope events are embedded within compressive MHD-timescale oscillations of large amplitude observed across all three magnetic field components, similar to the case on September 27.

After Event 4, we additionally identify an interval of approximately 50 seconds (indicated by the horizontal arrows at the top of Figure 5(a) and 5(b)), during which a group of closely packed magnetic field enhancements is observed. Unlike the four main events, the contributions from $B_N$ to these enhancements are much less prominent or not clearly discernible. Although a further decrease in $|V|$ is identifiable in at least two instances (vertical dark yellow arrows in Figure 5(c) and 5(d)) within this interval, the association between these multiple magnetic enhancements and plasma parameters is generally less distinct than in the four main events.

## 4. Discussion and Conclusion

We emphasize that the flux ropes are most clearly identifiable near the HCS center, where the background magnetic field is weakest and the relative enhancement in magnetic field strength within the flux rope becomes most prominent. This effect is particularly notable closer to the Sun, as in our cases which were found at heliocentric distances of ~12 $R_\odot$. At these distances, the ambient magnetic field is typically a few hundred nanotesla, making it more difficult for a flux rope to be distinguished from a higher B environment unless the spacecraft passes sufficiently close to the HCS center for a sufficient time duration. As these flux ropes are convected outward, they may expand due to the declining background pressure. In such regions where the surrounding magnetic field is even weaker, the likelihood of encountering these structures as distinct entities—potentially even at locations offset from the current sheet center—increases.

A relevant example to which this conjecture applies is a series of magnetic flux ropes reported by K.-E. Choi et al. (2024). They are characterized by enhanced B and lower density and temperature in a narrow Carrington longitudinal range observed by PSP at r~35-44$R_\odot$ - significantly farther from the Sun than our observation points near ~12$R_\odot$. Each of their flux ropes lasted from ~0.5 to 1.8 hours, much larger in scale than those observed in our study which lasted only seconds to tens of seconds. The authors attributed these observations to successive passage of flux ropes likely resulting from successive magnetic reconnection at one or more sites closer to the Sun than PSP. An important distinction between their observations and ours is that their flux ropes were not detected near the HCS center, without a HCS crossing by PSP, and no direct signatures of reconnection exhaust were found. However, we speculate that their flux ropes may have expanded substantially during outward propagation, allowing PSP to encounter them even when positioned some distance away from the HCS center.

A likely origin of the flux ropes observed within the reconnection exhaust is secondary reconnection driven by either outflow turbulence or plasmoid instability (W. H. Mattaeus & S. L. Lamkin 1986; J. F. Drake et al. 2006; S. Servidio et al. 2009; G. Lapenta et al. 2015; L. Comisso et al. 2016; C. Dong et al. 2018; H. Arro et al. 2020; H. Arnold et al. 2021; S. Eriksson et al. 2022; M. I. Desai et al. 2025) occurring within the HCS. For example, full particle simulations by J. F. Drake et al. (2006) show that when secondary reconnection develops under a guide field, the secondary islands grow to finite size before merging with the main magnetic island and form strong core fields. In the process of forming, secondary islands compress the ambient out-of-plane magnetic field and therefore evolve into flux tubes with strongly enhanced core fields. This feature is consistent with our observations, in which the flux ropes exhibit pronounced core field enhancements within the reconnection exhaust. Additionally, in a recent report, M. I. Desai et al. (2025) commented on large variations in the normal component of the magnetic field during an HCS crossing at ~16.25$R_\odot$, implying the presence of large-scale magnetic islands or flux ropes embedded within the reconnection exhaust—though no detailed analysis of the observed flux ropes was provided. This interpretation is also consistent with the structures observed in our two HCS crossing events. Moreover, simulations by

M. I. Desai et al. (2025) showed that the HCS can develop multiple reconnection sites, resulting in the formation of a large population of flux ropes that merge dynamically during the system's evolution. In their model, reconnection progresses through the initial formation of many small flux ropes, which subsequently coalesce into larger structures. Notably, the axial (guide) field component is found to increase significantly within the flux ropes relative to the upstream value, due to plasma compression in the reconnecting current layer. The simulation results align well with the key features observed in our events.

**Acknowledgement:**

This work was supported by the National Research Foundation of Korea(NRF) grant funded by the Korea government(MSIT)(RS-2024-00454886). We acknowledge the NASA Parker Solar Probe Mission and SWEAP team led by Justin Kasper for use of the data. The FIELDS experiment on the Parker Solar Probe spacecraft was designed and developed under NASA contract NNN06AA01C. SJN was supported by NASA contract no. NNG14PJ13I.